\documentclass[twocolumn,amssymb,amsmath,showpacs,nofootinbib,superscriptaddress]{revtex4}

\usepackage{graphicx}
\usepackage{dcolumn}
\usepackage{bm}
\usepackage{color,soul}

\begin{document}

\title{The Energy Dependence of Micro Black Hole Horizon in Quantum Gravity Theory}

\author{Lei Feng}
\email{fenglei@pmo.ac.cn}
\affiliation{Key Laboratory of Dark Matter and Space Astronomy, Purple Mountain Observatory, Chinese Academy of Sciences, Nanjing 210008, China}
\affiliation{College of Physics, Qingdao University, Qingdao 266071, China}

\begin{abstract}
The energy dependence of the deflection angle is a common prediction in some quantum gravity theories when the impact parameters are much larger than the photon wavelength. For low energy photons, the deflection angle recovers to the prediction of GR. But it reduces to zero for infinite energy photons. In this paper, we develop an effective approach to calculate the trajectory of photons and other deflection-related quantities semiclassically by replacing $h_{\mu \nu}$ with $h_{\mu \nu} \times f(E)$ to include the correction of quantum gravity. This approach could provide more information for photons traveling in an external gravitational field. We compute the horizon of micro black hole with this method and find that they are all energy dependent and decrease to zero as the energy increases to infinity.

\end{abstract}
\maketitle

\section{Introduction}

The general relativity (GR) is the most accurate gravity theory nowadays and it has been tested by different types of experimentes in the last one hundred years. The perihelion advance of Mercury is one of the earliest evidences. $\Delta \phi$ is about $43^{\prime\prime}$ per century in GR and the current measurement is $42.98\pm 0.4^{\prime\prime}$ per century \cite{perihelion}. Light bending, another important evidence, was firstly observed by Dyson and Eddington in 1919 and then confirmed with much better accuracy by the very long baseline radio interferometry\cite{lebach,beson}. The light travel time delay named Shapiro delay, which was firstly introduced by Shapiro \cite{Shapiro}, can also be used to test GR. The new measurements are consistent with the prediction of GR very well \cite{Rindler,ni}.

The gravitational lensing, as a result of the bending light, is also an important way to test GR which contains three main types: ¡®strong lensing¡¯, ¡®weak lensing¡¯ and ¡®microlensing¡¯. Lots of strong lensing system have already been observed by different experiments, such as the Lenses Structure and Dynamics survey \cite{Koopmans,Treu1,Treu2}, the Sloan Lens Advanced Camera
for Surveys \cite{Bolton, Auger}, the Baryon Oscillation Spectroscopic Survey \cite{Brownstein}, the Strong Lensing Legacy Survey \cite{Gavazzi,Sonnenfeld1,Sonnenfeld2} and the Dark Energy Survey \cite{DES}.

But GR, as one type of classical physics, is in conflict with the idea of quantum theory. GR can be quantized directly but it is not a renormalizable theory.
Then higher-derivative terms, such as $R^2, R^2_{\mu \nu},$, $R^2_{\mu \nu \alpha \beta}$ and so on, were introduced to solve such difficulties  \cite{Weyl,Eddington,Stelle}. Such theories are renormalizable but the annoying massive spin-2 ghost is unavoidable at the same time. Theorists then introduced the infinite derivative terms and built several kinds of nonlocal gravity models \cite{IDG1,IDG2,IDG3}. Such models can kill two birds with one stone which avoid the annoying missive ghost and divergence problem at small distance.

In this paper, we develop an effective approach to compute the trajectory of photons and other deflection-related quantities semiclassically in quantum gravity by redefining the perturbed metric $h_{\mu \nu}$. Our approach could provide more information on deflection-related issues. This method has obvious limitations especially when the impact parameters are much larger than the wavelength of photons. So we adopt with this method to compute the horizon of micro black hole in this draft.

This draft is organized as follows: In Sec. II we review the calculation of deflection angle
with classical and semiclassical method. In Sec. III, we present the new effective approach and our conclusions are summarized in Sec. IV.

\section{The Deflection Angle}
Light bending is an important prediction of GR and is about $1.75^{\prime\prime}$ for the photons gazing the sun.
The classical deflection angle is described by the following formula
\begin{eqnarray}
\theta_{\rm GR}= \frac{\kappa}{2} \int_{-\infty}^{\infty}{\partial_y[h_{00} + h_{11}]dx^1}.
\end{eqnarray}
Here we assume that the gravitational source and the trajectory of photons lie in $x-y$ plane. And the perturbed metric is defined as follows
\begin{equation}
g_{\mu \nu}= \eta_{\mu \nu} + h_{\mu \nu}.
\label{eq:2}
\end{equation}

\begin{figure}
\includegraphics[width=80mm,angle=0]{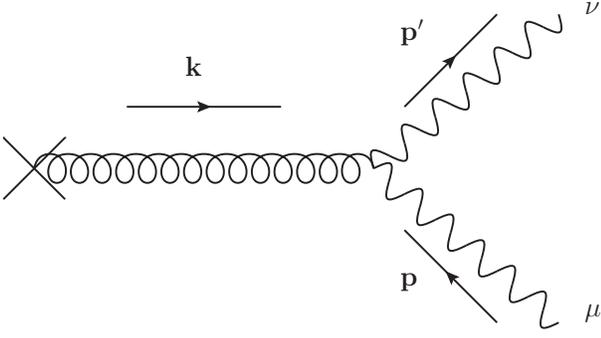}
  \caption{The Feynmann diagram of the interaction between external gravitational field and photon.}
  \label{fig:1}
\end{figure}

The gravitational deflection angle can also be calculated semiclassically within the framework of quantum gravity \cite{HDG1,HDG2,HDG3,Accioly1,Accioly2,Accioly3,Accioly4,Accioly5} which provides us much more information.
The Feynman diagram of photons scattering in the external gravity field is shown in Fig. \ref{fig:1} and the corresponding amplitude is

\begin{eqnarray}
{\cal{M}}_{r r'} =&& \frac{1}{2} \kappa h^{\lambda \rho}_{\mathrm{ext}}({\bf{k}})\Bigg[-\eta_{\mu \nu} \eta_{\lambda \rho}pp' + \eta_{\lambda \rho}p'_\mu p_\nu + 2\Big(\eta_{\mu \nu}p_\lambda p'_\rho  \nonumber \\  &&- \eta_{\nu \rho}p_\lambda p'_\mu  \nonumber - \eta_{\mu \lambda }p_\nu p'_\rho + \eta_{\mu \lambda} \eta_{\nu \rho}pp'\Big) \Bigg]\epsilon^\mu_r({\bf{p}}) \epsilon^\nu_{r'}({\bf{p'}}) \nonumber,
\end{eqnarray}

\noindent where $\epsilon^\mu_r({\bf{p}})$ and $\epsilon^\nu_{r'}({\bf{p'}})$ are the polarization vectors of photons and they satisfies the following summation relation

\begin{equation}
\sum_{r=1}^{2}\epsilon^\mu_r({\bf{p}})\epsilon^\nu_{r}({\bf{p}})= -\eta^{\mu \nu} - \frac{p^\mu p^\nu}{(p\cdot n)^2} + \frac{p^\mu n^\nu + p^\nu n^\mu}{p\cdot n},
\end{equation}

\noindent where $n^2=1$.
$h^{\lambda \rho}_{\mathrm{ext}}({\bf{k}})$ denotes the gravitational field in momentum space. It has different formula for specific quantum gravity theory and it is defined as follows

\begin{equation}
h^{\lambda \rho}_{\mathrm{ext}}({\bf{k}})= \int{d^3{\bf{r}} e^{-i{\bf{k}}\cdot {\bf{r}}}h^{\lambda \rho}_{\mathrm{ext}}({\bf{r}})}.
\label{eq:k}
\end{equation}

Here we briefly review the result of nonlocal gravity as an example \cite{IDG-FENG}. The gravitational action for nonlocal gravity is expressed as
\begin{equation}
S=-\frac{1}{16\pi G}\int d^4x \sqrt{-g} \{R+ G_{\mu\nu}\frac{a(\square)-1}{\square}R^{\mu\nu}\},
\label{eq:s}
\end{equation}
where $G_{\mu\nu}=R_{\mu\nu}-\frac{1}{2}g_{\mu\nu}R$ is the famous Einstein matrix and $a(\square)=e^{-\frac{\square}{\Lambda^2}}$. $\Lambda$ is the energy scale non-locality and the current constrain on energy scale $\Lambda$ \cite{lowerlimit,lowerlimit2} is
$\rm \Lambda > 0.01~eV$.

For nonlocal gravity, $h^{\lambda \rho}_{\mathrm{ext}}({\bf{k}})$ is presented as
\begin{equation}
h^{{\mathrm{(E)}} \mu \nu}_{\mathrm{ext}}({\bf{k}}) = \kappa  M\left(\frac{\eta^{\mu \nu}}{2{\bf{k}}^2} - \frac{ \eta^{\mu 0} \eta^{\nu 0}}{\bf{\bf{k}}^2}\right){\rm exp(-\frac{{\bf{k}}^2}{\Lambda^2}}).
\end{equation}

Then we can easily get the unpolarized cross-section, which is
\begin{eqnarray}
\frac{d\sigma}{d \Omega}&&= \frac{1}{(4 \pi)^2} \frac{1}{2}\sum_{r} \sum_{r'}{\cal{M}}^2_{r r'} \nonumber \\ &&= \frac{1}{(4 \pi)^2 }\frac{\kappa^4 M^2 E^4(1 + \cos{\theta})^2}{16} \Bigg[\frac{1}{{\bf{k}}^2}{\rm exp(-\frac{{\bf{k}}^2}{\Lambda^2})} \Bigg]^2, \nonumber
\end{eqnarray}
where $E$ is the energy of the injected photon and $\theta $ denotes the angle between $\bf p$ and $\bf p^\prime$.

After careful calculations in the small angle case, we finally got the energy dependence of deflection angle \cite{IDG-FENG} which is governed by the following equation
\begin{eqnarray}
\frac{1}{\theta^2_{\mathrm{GR}}}= \frac{1}{\theta^2} exp(-\frac{2\theta^2}{\lambda^2}) +  \frac{2}{\lambda^2}Ei(-\frac{2\theta^2}{\lambda^2}),
\label{eq:a}
\end{eqnarray}
where $\lambda \equiv \frac{\Lambda}{E}$ and $\theta$ is the deflection angle.  The exponential integral $Ei(x)$ is defined as follows
\begin{eqnarray}
Ei(x)=-\int_{-x}^{\infty}\frac{e^{-t}}{t}dt.
\end{eqnarray}

The deflection angle is derived by solving the above equation numerically \cite{IDG-FENG} which is energy dependent. If $\Lambda/E \rightarrow \infty$, Eq. (\ref{eq:a}) becomes $\theta=\theta_{\rm GC}$. In other words, it recovers the prediction of GR for low energy photons. The second term is negative in the right-side of Eq. (\ref{eq:a}) and then first term must be positive. So if $\Lambda/E \rightarrow 0$, $\theta$ should also approach to 0. This means that there is no deflection for photons with sufficiently high energy.

In\cite{HDG1,HDG2,HDG3}, the authors calculated the gravitational deflection angle in HDG model and found that it decreases to zero at $log_{10}|\beta|\sim 89$. In Ref. \cite{xu2017}, the authors systematical studied the deflection angle in several quantum gravitational theories and found that it is an common conclusion.

It should be noted that that this semiclassical approach based on Feynman diagrams
is not efficient for impact parameters which are much larger than the photon wavelength.

\section{The Effective Approach}

Here we develop an effective way to deal with the photon deflection problems in quantum gravity theories. Firstly, we introduce a dimensionless function $f(E)$ which is
\begin{eqnarray}
f(E,b)=\frac{\theta(E)}{\theta_{\rm GR}(b)},
\label{eq:fe}
\end{eqnarray}
where b is the impact parameter. From the above definition, we can see that the form of $f(E)$ depends on the specific quantum gravity theory and it has the following behavior
\begin{eqnarray}
f(E,b)\rightarrow &1&~if~E\rightarrow 0,  \nonumber \\
\\
f(E,b)\rightarrow &0&~if~E\rightarrow \infty. \nonumber
\label{eq:fe2}
\end{eqnarray}
Then we define the following effective metric in cartesian coordinates:
\begin{eqnarray}
h^\prime_{\mu \nu}=h_{\mu \nu} \times f(E,b).
\label{eq:hprime}
\end{eqnarray}
Substituting Eq.(\ref{eq:hprime}) into Eq.(\ref{eq:2}), we can conveniently reproduce the deflection angle of quantum gravity.

The result that $f(E,b)\rightarrow {\rm 0~ when~E\rightarrow \infty}$ is gotten in the weak gravity situation with semiclassical approach and here we assume that our effective approach is still valid for the system of micro black holes.

Here, we adopt with this method to compute the horizon of micro black hole. In principle, the smallest mass of the micro black hole is approximately the Planck mass. Correspondingly, the smallest horizon is $R_h=2GM_p/c^2~\sim 10^{-33}~cm$, where $M_p$ is the Planck mass and c is the speed of light.
Therefore, the horizon and the impact parameters could be small enough comparing with the photon wavelength.

In this draft, we only discuss the situation of micro Schwarzschild black hole for simplicity. The Schwarzschild metric is described by  the following formula
\begin{eqnarray}
ds^2=&-&(1-2M/r)dt^2+(1-2M/r)^{-1}dr^2 \nonumber  \\
    &+&r^2(d\theta^2+\sin^2\theta d \phi^2),
\label{eq:schw}
\end{eqnarray}
where M is the mass of Schwarzschild black hole. Doing the coordinate transformation as follows:
\begin{eqnarray}
r&=&r^\prime(1+M/2r^\prime),\nonumber  \\
r^\prime&=&\frac{1}{2}(\sqrt{r^2-2Mr}+r-M), \nonumber  \\
x&=& r^\prime \sin(\theta)\cos(\phi),\\
y&=& r^\prime \sin(\theta)\sin(\phi), \nonumber  \\
z&=& r^\prime \cos(\theta), \nonumber
\label{eq:coordinate-transformation}
\end{eqnarray}
we can get the corresponding cartesian coordinates form of Schwarzschild metric, which is
\begin{eqnarray}
ds^2=&-&[\frac{1-M/2r^\prime}{1+M/2r^\prime}]^2dt^2 \nonumber  \\
&+&(1+M/2r^\prime)^4[dx^2+dy^2+dz^2].
\label{eq:kerrcartesian}
\end{eqnarray}
Then the replacement for the space component $h_{xx,yy,zz}$ is $f(E,b)[4M/2r^\prime+6(M/2r^\prime)^2+4(M/2r^\prime)^3+(M/2r^\prime)^4]$.
Then we let
\begin{eqnarray}
1+f^\prime(E,b,r)&=&1+f(E,b)[4M/2r^\prime+6(M/2r^\prime)^2 \nonumber \\
&+& 4(M/2r^\prime)^3+(M/2r^\prime)^4],
\label{eq:replacement2}
\end{eqnarray}
and $f^\prime(E,b,r)\rightarrow 0$ when $E\rightarrow \infty$ and we write $f^\prime(E,b,r)$ as $f^\prime(E)$ for short in the rest of this paper.
Converting to the original polar coordinate form Eq. (\ref{eq:schw}), the coefficient of $dr^2$ is
\begin{eqnarray}
[1+f^\prime(E)][\frac{[\frac{1}{2}(\sqrt{r^2-2Mr}+r-M)]^2}{r(r-2M)}]dr^2.
\label{eq:kerr-corrected3}
\end{eqnarray}

The quantum-corrected black hole horizon can be deviated by the light-like hypersurface condition, which is
\begin{eqnarray}
\frac{1}{1+f^\prime(E)}\frac{r(r-2M)}{[\frac{1}{2}(\sqrt{r^2-2Mr}+r-M)]^2}=0.
\label{eq:kerr-corrected}
\end{eqnarray}
When $E\rightarrow \infty$, $f^\prime(E)\rightarrow 0$ \emph{}and the solutions of the above equation are ${\rm 0~and~2M}$.
Obviously, $\rm r=0$ is a new solution for light-like hypersurface condition because of the extra term $f^\prime(E)\rightarrow 0$.
This means that the horizon of micro black hole decrease to zero for infinite energy photons. This is a reasonable result of quantum gravity because there is no deflection for photons with high enough energy.


From the above analysis, it is obvious that the horizon of micro Schwarzschild black hole is energy dependent and tends to zero for photons with infinitely energy.
For the micro Kerr and Kerr-Newman black hole, the calculation processes are much more complicated and the corresponding horizon might be similar with the situation of the micro Schwarzschild black hole.

\section{Summary}

In this draft, we developed an effective method to calculate the deflection-related quantities. Replacing $h_{\mu \nu}$ with $h_{\mu \nu} \times f(E)$ is the main idea of our approach in order to include the correction of quantum gravity. By this approach, we can easily calculate the trajectory and other deflection-related quantities for photons with small enough impact parameters. From our analysis we found that the horizon of micro black hole is energy dependent and decreases to zero as energy increases to infinity.

The current conclusion may be not applicable for normal black hole because the semiclassical approach is invalid for large impact parameters.  Considering the strong gravity near the black hole, the quantum effect might be not ignorable. The horizon of normal black might be very complex taking into account the gravitational UV approaching behavior.

The energy dependence of the micro black hole horizon is an unexpected result. It may conflict with the currently black hole thermodynamics theory because the size of black hole is also energy dependent. The current definition of the black hole entropy might be also unsuitable for micro black hole.

\acknowledgments
We thank Dr. Yuan-Yuan Chen for helpful discussions and suggestions. This work was supported in part by  the National Key Program for Research and Development (2016YFA0400200),  the National Natural Science of China (Nos. 11773075, U1738206) and the Youth Innovation Promotion Association CAS (Grant No. 2016288).
\\


\begin{thebibliography}{}

\bibitem{perihelion} Lo, K.-H., Young, K. \& Lee, B.Y.P., 2013. Advance of perihelion. American journal of physics, 81(9), p.695.
\bibitem{lebach}{D. Lebach  {\it  et al.},  Phys. Rev. Lett. \textbf{75}, 1439 (1995).}
\bibitem{beson}{E. Fomalont, S. Kopeikin, G. Lanyi, and J. Benson,  Astrophys. J. \textbf{699}, 1395 (2009).}
\bibitem{Shapiro} Shapiro, I.I., 1964. Fourth Test of General Relativity. Phys. Rev. Lett., p.789.
\bibitem{Rindler} Rindler, W., 2006. Relativity: Special, general, and cosmological 2nd ed., Oxford: Oxford University Press.
\bibitem{ni} Ni, W.T., 2016. Solar-system tests of the relativistic gravity. arXiv:1611.06025.
\bibitem{Koopmans}Koopmans L. V. E., Treu T., 2003, ApJ, 583, 137
\bibitem{Treu1} Treu T., Koopmans L. V. E., 2002, ApJ, 575,87
\bibitem{Treu2} Treu T., Koopmans L. V. E., 2004, ApJ, 611, 739
\bibitem{Bolton} Bolton A. S., Burles S., Koopmans L. V. E., Treu T., Gavazzi R., Moustakas L. A., Wayth R.,
Schlegel D. J., 2008, ApJ, 682, 964
\bibitem{Auger} Auger M. W., Treu T., Bolton A. S., Gavazzi R., Koopmans L. V. E., Marshall P. J., Bundy K.,
Moustakas L. A., 2009, ApJ, 705, 1099
\bibitem{Brownstein} Brownstein J. R. et al. 2012, Apj, 744, 41

\bibitem{Gavazzi} Gavazzi R., Marshall P. J., Treu T., Sonnenfeld A., 2014, ApJ, 785, 144
\bibitem{Sonnenfeld1} Sonnenfeld A., Gavazzi R., Suyu S. H., Treu T., Marshall P. J., 2013, ApJ, 777, 97
\bibitem{Sonnenfeld2} Sonnenfeld A., Treu T., Gavazzi R., Suyu S. H., Marshall P. J., Auger M. W., Nipoti C., 2013,
ApJ, 777, 98
\bibitem{DES} Diehl H. T. et al., 2017, ApJS, 232, 15


\bibitem{Weyl}{H. Weyl, {\it Space-Time Matter} (Dover, 1952).}
\bibitem{Eddington}{ A. Eddington, {\it The Mathematical Theory of Relativity}, 2nd. ed.  (Cambridge University Press,  1924).}
\bibitem{Stelle}{K. Stelle, Phys. Rev. D \textbf{16}, 953 (1977).}

\bibitem{IDG1}L. Modesto, Super -renormalizable Quantum Gravity ,
Phys. Rev. D \textbf{86}, 044005, [arXiv:1107.2403v1 [hep-th]].
\bibitem{IDG2} T. Biswas, E. Gerwick, T. Koivisto and A. Mazumdar,
Towards singularity and ghost free theories of gravity,
Phys. Rev. Lett. \textbf{108}, 031101, [arXiv:1110.5249 [grqc]].

\bibitem{IDG3} L. Modesto, Super -renormalizable Higher -Derivative
Quantum gravity, [arXiv:1202.0008v1 [hep-th]]. L.
Modesto, T. de Paula Netto and Ilya. L. Shapiro, On
Newtonian singularities in higher derivative gravity models,
High Energ. Phys. (2015) \textbf{2015}: 98. [arXiv:1412.0740
[hep-th]].

\bibitem{HDG1} A. Accioly, J. Helay$\rm \ddot{e}$l-Neto, B. Giacchini and W. Herdy, Phys.Rev. D\textbf{91} (2015) no.12, 125009
\bibitem{HDG2} A. Accioly et. al.,  arXiv:1604.07348
\bibitem{HDG3} A. Accioly, B. L. Giacchini and I. L. Shapiro, arXiv:1610.05260

\bibitem{Accioly1}{ R. Paszko and A. Accioly, Class. Quantum Grav. { \bf 27}, 145012 (2010).}

\bibitem{Accioly2}{A. Accioly and R. Paszko, Int. J. Mod. Phys. D  { \bf 18}, 2107 (2009).}

\bibitem{Accioly3}{A. Accioly  and R. Paszko,  Adv. Stud. Theor. Phys.  { \bf 3}, 65 (2009).}

\bibitem{Accioly4}{A. Accioly and R. Paszko, Phys. Rev. D  { \bf 78}, 064002 (2008).}
\bibitem{Accioly5}{A. Accioly, R. Aldrovandi, and R. Paszko, Int. J. Mod. Phys. D  { \bf 15}, 2249 (2006).}
\bibitem{IDG-FENG} Lei Feng, Phys.Rev. D95 (2017) no.8, 084015

\bibitem{lowerlimit} J. Edholm, A. Koshelev and A. Mazumdar, Universality
of testing ghost -free gravity, 2016, [arXiv:1604.01989 [grqc]].
\bibitem{lowerlimit2} D. J. Kapner, T. S. Cook, E. G. Adelberger, J. H. Gundlach, B. R. Heckel, C. D.
Hoyle and H. E. Swanson, Phys. Rev. Lett. \textbf{98} (2007) 021101.
\bibitem{xu2017} Chenmei Xu and Yisong Yang, arXiv:1709.04127





\end{thebibliography}
\end{document}